\documentclass[10pt]{article} 
\usepackage{graphics} 
\usepackage{epsfig} 

\title{Off-Axis Neutrino Scattering\\ in GRB Central Engines} 

\author{ Nathan D. George\thanks{Department of Mathematics, North
Carolina State University, Raleigh, NC 27695--8205}\ \and Arkady
Kheyfets${}^{*}$\ \and John M. McGhee\thanks{Computer and
Computational Sciences Division, Los Alamos National Laboratory, Los
Alamos, NM 87545}\ \and Warner A.~Miller\thanks{Theoretical
Astrophysics Group (T-6, MS B288), Theoretical Division, Los Alamos
National Laboratory, Los Alamos, NM,87544, and Air Force Weapons
Laboratory, High Power Microwave Division, Kirtland AFB,
NM 87117; Manuscript Contact E-mail: wam@lanl.gov} }

\date{\today} 

\newcommand{\bu}{\bar{\mathbf{u}}} 
\newcommand{\pnu}{\mathbf{p}_\nu} 
\newcommand{\pbnu}{\mathbf{p}_{\bar\nu}} 
\newcommand{\Omnu}{\mathbf{\Omega}_\nu} 
\newcommand{\Ombnu}{\mathbf{\Omega}_{\bar\nu}} 
\newcommand{\be}{\mathbf{e}} \newcommand{\fnu}{f_\nu} 
\newcommand{\fbnu}{f_{\bar\nu}} \newcommand{\epnu} {\varepsilon_\nu} 
\newcommand{\epbnu} {\varepsilon_{\bar\nu}} 
\newcommand{\vnu}{\mathbf{v}_\nu} 
\newcommand{\vbnu}{\mathbf{v}_{\bar\nu}} \newcommand{\bpp}{\mathbf{p}} 
\newcommand{\bppf}{\bar{\mathbf{p}}} 
\newcommand{\ppr}{\mathbf{p}_{\parallel\nu}}

\begin{document} 
\maketitle 

\begin{abstract} 
  The search for an understanding of an energy source great enough to
  explain the gamma-ray burst (GRB) phenomena has attracted much
  attention from the astrophysical community since its discovery.  In
  this paper we extend the work of K. Asano and T. Fukuyama, and J. D.
  Salmonson and J. R. Wilson, and analyze the off-axis contributions
  to the energy-momentum deposition rate (MDR) from the $\nu
  \bar{\nu}$ collisions above a rotating black hole/thin accretion
  disk system. Our calculations are performed by imaging the accretion
  disk at a specified observer using the full geodesic equations, and
  calculating the cumulative MDR from the scattering of all pairs of
  neutrinos and anti-neutrinos arriving at the observer.  Our results
  shed light on the beaming efficiency of GRB models of this kind.
  Although we confirm Asano and Fukuyama's conjecture as to the
  constancy of the beaming for small angles away from the axis;
  nevertheless, we find the dominant contribution to the MDR comes
  from near the surface of the disk with a tilt of approximately
  $\pi/4$ in the direction of the disk's rotation.  We find that the
  MDR at large radii is directed outward in a conic section centered
  around the symmetry axis and is larger, by a factor of 10 to 20,
  than the on-axis values. By including this off-axis disk source, we
  find a linear dependence of the MDR on the black hole angular
  momentum ($a$).  In addition, we find that scattering is directed
  back onto the black hole in regions just above the horizon of the
  black hole. This gravitational ``in scatter'' may provide an
  observable high energy signature of the central engine, or at least
  another channel for accretion.

\end{abstract} 

\section{Neutrino Scattering About a Low Mass Accreting Kerr Black Hole} 
\label{sec0} 

The search for an understanding of an energy source great enough to
explain the gamma-ray burst (GRB) phenomena has attracted much
attention from the astrophysical community since its discovery.  One
of the more recent candidates for the GRB model is based on the
fire-ball model.\cite{Mez2002,CavRee1978,Pac1986,ShePir1990} The
proposal for the central engine for such bursts concerns $\nu
\bar{\nu}$ production and their subsequent scattering in the environs
of a dense accretion disk/low-mass black hole system \cite{Woo1993,
  PopWooFry1999}.  Such a system may be formed by the merger of a
black hole and a neutron star, the merger of two neutron stars, by a
collapsar, or by a supernova
\cite{Woo1993,EicLivPirSch1989,MacWoo1998,ViePerPirSte2000}.  The
astrophysical details of the geometry or environment of such systems
are currently hidden from us both observationally and computationally,
although this situation may change in the near future with the
development of gravity wave observatories and more sophisticated
general relativistic astrophysical simulation capabilities. For the
mean time in this paper we focus our attention on a rather simplified
model for the scattering of neutrinos and anti-neutrinos in the
vicinity of a low-mass rotating black hole.  Our goal is to extend the
recent work of Asano \& Fukuyama (AF) \cite{AsaFuk2001,AsaFuk2000} and
Salmonson and Wilson \cite{SalWil1999} (SW) by providing the first
general-relativistically covariant off-axis calculation of the energy
momentum deposition rate (MDR). We focus our attention, as did the
previous authors, on the general relativistic enhancement or
degradation of the deposited MDR.

Since we now intend to calculate the off-axis contributions, the direction 
of the energy-momentum 4-vector can no longer be assumed, as opposed 
to the along axis case, where symmetry forces the net energy upward 
along the symmetry axis. An accurate treatment of off-axis contributions 
requires a full 4-dimensional spacetime 
calculation, as opposed to the analytic and semi-analytic calculations 
performed by AF and SW.  Therefore, we derive the full energy-momentum 
(hereafter referred to as "momenergy")\cite{Whe1990} deposited in the 
vicinity of the black hole/accretion disk system.  The calculations 
are based on the basic principles as discussed in Misner, Thorne and 
Wheeler.\cite{MisThoWhe1972} We consider the direction of the 
momenergy in the frame of an observer at a given point above the 
disk. 

Our aim in this work is not solely to study and draw conclusions on
the relativistic case of neutrino and anti-neutrino collisions above
the accretion disk for GRB models, but also to develop a method to
provide a more in-depth understanding of the scattering mechanisms in
the environs of small accreting black hole systems.  For the above
purpose, we consider an idealized situation which includes
gravitational redshift, the bending of neutrino trajectories, and the
redshift due to accretion disk rotation.

This paper is organized as follows: In Sec.~\ref{sec1} and 
Sec.~\ref{sec2} we define the appropriate global spacetime metric and 
establish the frame of the observer.  The derivation of the momenergy 
deposition rate (MDR) is discussed in Sec.~\ref{sec3}. 
Sec.~\ref{sec4} deals with the computation of the MDR in an 
observer's frame with regards to evaluating the number densities and 
the generation of the 4-momentum from the $\nu$-$\bar\nu$ scattering. 
In Sec.~\ref{sec5} we give an estimate of the cumulative effect 
of the scattering on the energy deposition at a radius of $50M$ 
from the black hole.  Finally, in Sec.~\ref{sec6} we draw some 
conclusions, discuss the applicability of our results and methods on 
a larger scale, and make recommendations for future work in this area. 

\section{The Global Spacetime Metric} 
\label{sec1} 
        
We use the Boyer-Lindquist (BL) spacetime metric in our neutrino ray 
tracing code. 
\begin{equation} 
ds^2 = g_{tt}\, dt^2 + 2\, g_{t\phi}\, dt\, d\phi + g_{rr}\, dr^2 + 
g_{\theta\theta}\, d\theta^2 + g_{\phi\phi}\, d\phi^2 
\end{equation}     
where 
\begin{eqnarray} 
g_{tt} & = & -1 + \frac{2 M r}{\Sigma}\\ 
g_{t\phi} & = & - \frac{2 M r}{\Sigma} a \sin^2\theta\\ 
g_{rr} & = & \frac{\Sigma}{\Delta}\\ 
g_{\theta\theta} & = & \Sigma\\ 
g_{\phi\phi} & = &  \left(\frac{2 M r}{\Sigma}\, a^2 \sin^2\theta + r^2 + 
a^2\right) \sin^2\theta\\ 
\Sigma & = & r^2 + a^2 \cos\theta\\ 
\Delta & = & r^2 + a^2 - 2 M r 
\end{eqnarray}       
with $0 \le a \le M$. 

\section{The Frame of the Observer} 
\label{sec2} 

The momenergy deposition rate (MDR) is computed in the frame of an
observer resting with respect to the global coordinates $(t, r,
\theta, \phi )$ at a point with coordinates $r, \theta, \phi =
const$. Spacetime is stationary which implies that spatial pictures
are sufficient. Also, $\phi$ can be fixed (say $\phi = 0$); however,
the observational effects are nonetheless ordinarily only
representable in 3-dimensions because of frame dragging.

The frame of the observer is defined by four orthonormal basis vectors
($\{\be_{\hat\mu}\}_{{\hat\mu} = 0}^3$). We define these basis vectors
such that (1) the basis is a Lorentz frame ($\be_{\hat\mu} \cdot
\be_{\hat\lambda} = \eta_{\hat\mu\hat\lambda}$), and (2) the
4-velocity, or temporal basis vector of the observer ($\bu =
\be_{\hat0}$) is parallel to the time component of the
BL coordinate frame, $\be_0\equiv\partial_t$.

In the global coordinate frame 
\begin{equation} 
\be_{\hat 0} = \alpha \partial_t 
\end{equation} 
\begin{equation} 
\be_{\hat 0} \cdot \be_{\hat 0} = \alpha^2 \partial_t \cdot \partial_t = g_{00} \alpha^2 = 
-1 \Longrightarrow \alpha = \frac{1}{\sqrt{- g_{00}}} 
\end{equation} 
This leads to 
\begin{equation} 
\label{e0h} 
\be_{\hat 0} = \frac{1}{\sqrt{-g_{00}}}\, \partial_t 
\end{equation} 
Likewise 
\begin{eqnarray} 
\be_{\hat r} & = & \frac{1}{\sqrt{g_{rr}}}\, \partial_r \label{erh}\\   
\be_{\hat \theta} & = & \frac{1}{\sqrt{g_{\theta\theta}}}\, 
\partial_\theta \label{eth} 
\end{eqnarray} 
Vector $\be_{\hat \phi}$ is determined by the conditions 
$\be_{\hat \phi} = \alpha\, \partial_t + \beta\, \partial_\phi$, 
$\be_{\hat \phi} \cdot  \partial_t = 0$, $\be_{\hat \phi} \cdot 
\be_{\hat \phi} = 1$, which leads to 
\begin{equation} 
\label{eph} 
\be_{\hat \phi} = \sqrt{\frac{g_{tt}}{g_{\phi\phi} g_{tt} - g_{t\phi}^2}} 
\left( -\frac{g_{t\phi}}{g_{tt}}\, \partial_t + \partial_\phi\right) 
\end{equation} 

\section{The Momenergy Deposition Rate in the Frame of the Observer} 
\label{sec3} 

The MDR in the frame of the observer is given by 
\begin{equation} 
\label{MDR} 
MDR = \int\limits_\nu\int\limits_{\bar\nu} 
\fnu (\pnu , \ldots )\, \fbnu (\pbnu , \ldots )\, 
\{ \sigma \vert \vnu - \vbnu\vert \epnu \epbnu\}\,   
(\pnu + \pbnu)\, \frac{d^3\pnu}{\epnu}\, \frac{d^3\pbnu}{\epbnu} 
\end{equation} 
where $\fnu$, $\fbnu$ are number densities in phase space (Lorentz invariant), 
and $\sigma$ is the rest frame cross section. The expression in curly brackets 
is Lorentz invariant and can be computed as 
\begin{equation}   
\{ \sigma \vert \vnu - \vbnu\vert \epnu \epbnu\} = 
\frac{D G_F^2}{3\pi} \left( -\epnu \epbnu + \pnu \cdot \pbnu\right)^2 
\end{equation} 
where 
\begin{equation} 
G_F^2 = 5.29 \times 10^{-44} cm^2\, MeV^{-2} 
\end{equation} 
\begin{equation} 
\label{DW} 
D = 1 \pm 4 \sin^2\theta_W + 8 \sin^4\theta_W 
\end{equation} 
In Eq.~(\ref{DW}), $\theta_W$ is the Weinberg angle, 
$\sin^2\theta_W = 0.23$ and the plus sign is used for $\nu_e\, \bar\nu_e$ 
pairs, while minus is used for $\nu_\mu\, \bar\nu_\mu$ and 
$\nu_\tau\, \bar\nu_\tau$ pairs. 

In order to perform integration in spherical coordinates, we 
use expressions $\pnu = \epnu\, \Omnu$ and 
$d^3\pnu = \epnu^2 d\epnu d\Omega_\nu$, where $\Omnu$ is the unit vector 
in the 
direction of $\pnu$ and $d\Omega_\nu$ is the solid angle element. Equation 
(\ref{MDR}) then reduces to 
\begin{equation} 
\label{MDR1} 
MDR = \frac{D G_F^2}{3\pi} \int\limits_\nu\int\limits_{\bar\nu} \fnu \fbnu 
(\Omnu \cdot \Ombnu - 1)^2 (\epnu\, \Omnu + \epbnu\, \Ombnu )\, \epnu^3 \epbnu^3\, 
d\epnu\, d\epbnu\, d\Omega_\nu\, d\Omega_{\bar\nu} 
\end{equation}

\section{Computing the Number Densities $\fnu$, $\fbnu$ } 
\label{sec4}

\begin{figure}[ht] 
\begin{center} 
\epsfig{file=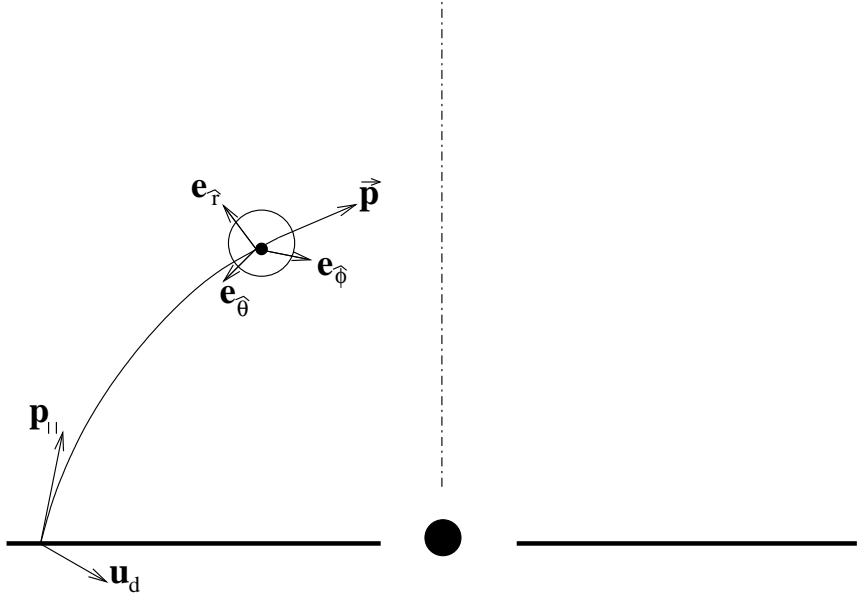,width=1.0\textwidth} \end{center} 
\caption{The geometry behind the computation of the total momenergy 
deposition rate in the observer's orthonormal frame (hatted basis 
vectors).} 
\end{figure}

The number densities $\fnu$, $\fbnu$ are Lorentz invariant and conserved 
along the $\nu$, $\bar\nu$ world lines (null geodesics). If $\bppf = 
\langle\varepsilon , \bpp\rangle$ is the neutrino (anti-neutrino) 4--momentum 
in the observer frame ($\varepsilon = \vert \bpp \vert$),  and 
$\bppf_{\parallel} = \langle\varepsilon_{\parallel}, \bpp_{\parallel}\rangle$ 
is the 4--momentum of the same neutrino in the comoving frame of the disk 
at the point of emission ($\varepsilon_{\parallel} = 
-\bu_e\cdot\bppf_{\parallel}$) then $\fnu$ is computed as 
\begin{equation} 
\label{FNU} 
\fnu = \fnu (\pnu , \ldots ) = \frac{I_\nu}{\nu^3}\left(\frac{1}{h^4}\right) = 
\frac{2}{h^3}\left({\mathrm{e}}^{\varepsilon_{\parallel\nu}/kT} + 1\right)^{-1}\end{equation} 
where $I_\nu$ is the specific intensity of radiation at a given 
frequency $\nu$.  Note that the ratio $\frac{I_\nu}{\nu^3}$ is Lorentz 
invariant by Liouville's Theorem.  The value of $\varepsilon_{\parallel\nu}$ in this equation is determined by the 4--momentum of the neutrino parallel translated to the disk along the world line of the neutrino $\ppr$ and on the 4--velocity of the particle in the disk emitting the neutrino $\bu_e$. The latter is computed as 
\begin{equation} 
\bu_d =\langle u^t, 0, 0 ,u^\phi\rangle = 
\left\langle\frac{dt}{d\tau}, 0, 0, \frac{d\phi}{d\tau}\right\rangle = 
\frac{dt}{d\tau}\, \left\langle 1, 0, 0, \frac{d\phi}{dt}\right\rangle 
\end{equation} 
where\cite{Bar1972} 
\begin{equation} 
\frac{d\phi}{dt} = \Omega = \frac{M^{1/2}}{r^{3/2} + a M^{1/2}} 
\end{equation} 
for the direct circular orbit of the emitting particle.

The normalization condition 
\begin{equation} 
\bu_d \cdot \bu_d = -1 
\end{equation} 
written as 
\begin{equation} 
\left(\frac{dt}{d\tau}\right)^2 \left( g_{tt} + 2 g_{t\phi}\, \Omega + 
g_{\phi\phi}\, \Omega^2\right) = -1 
\end{equation} 
allows the evaluation of $u^t$ as 
\begin{equation} 
u^t = \frac{dt}{d\tau} = \left[ -\left( g_{tt} + 2 g_{t\phi}\, \Omega + 
g_{\phi\phi}\, \Omega^2\right)\right]^{-\frac{1}{2}} 
\end{equation} 
The value of $\bu_d$ is given by 
\begin{equation} 
\bu_d = \left\langle u^t, 0, 0, u^t\, \frac{d\phi}{dt}\right\rangle 
\end{equation}

The steps for evaluating $\fnu (\vec p)$ are as follows: 
\begin{enumerate} 

\item $\vec p \longrightarrow \pnu = \langle\vert\vec p\vert , \vec p\rangle$ 
in the frame of observer (supplied by the integrating program) 

\item $\pnu = \langle p^t, p^r, p^{\theta}, p^\phi\rangle$ Transformation to 
global coordinates 

\item Tracing geodesics starting at $(t, r, \theta ,\phi )_{obs}$ tangent to 
$\pnu$ to the disk. Pick the geodesic parameter $\lambda$ such that 
$\frac{d}{d\lambda} = \pnu$, integrate back to $\theta = \frac{\pi}{2}$ 
($\lambda = 0$ at the observer, $\lambda < 0$ at the disk). 
This procedure 
yields $(t, r, \theta, \phi )_{disk}$ and $\ppr = (d/d\lambda )_{disk} = 
\langle p^t, p^r, p^{\theta}, p^\phi\rangle_{disk}$   

\item Computing $\bu (t, r, \theta , \phi )_{disk} = \bu_d$   

\item Computing $\varepsilon_\parallel = -\bu_d \cdot \pnu$ 

\item Computing $f(\vec p ) = f(\varepsilon_\parallel )$ using equation 
(\ref{FNU}) 

\end{enumerate}   

Assuming the emission rates of the neutrinos and the anti-neutrinos at 
the disk are identical, and assuming the neutrinos are emitted 
isotropically as a Planck black-body spectrum characterized by the 
temperature $T_{disk}$ in the frame comoving with the disk material, then 
the spectrum of neutrinos that our observer sees is also a Planck 
spectrum.   However, the temperature characterizing our observer's 
spectrum is scaled by the frequency shift of the neutrinos. 
\begin{equation} 
\label{Tscale} 
T_{\cal O} = \frac{\varepsilon_{\cal O}}{\varepsilon_\parallel} T_{disk} 
\end{equation} 
Following the derivation of Salmonson and Wilson\cite{SalWil1999},
the MDR integral (Eq.\ref{MDR1}) can be transformed into a
4-dimensional integral by analytically integrating over
$\varepsilon_{\nu}$ and $\varepsilon_{\bar{\nu}}$.
\begin{equation} 
\label{MDR2} 
MDR = \frac{7 \pi^3 \zeta(5) D G_F^2}{2 h^6} 
\int\limits_\nu\int\limits_{\bar\nu} 
(k {{\bar T}_{\cal O}})^4 (kT_{\cal O})^5   
(\Omnu \cdot \Ombnu - 1)^2 \Omnu\, d\Omega_\nu\, d\Omega_{\bar\nu} 
\end{equation} 
Here, the temperature ${T}_{\cal O}$ is the temperature of the black 
body spectrum of the neutrinos as measured by the observer ${\cal 
O}$. It will ordinarily depend on both the sky angles ($\psi$, $\xi$) we 
integrate over. In a similar way ${\bar T}_{\cal O} = {\bar T}_{\cal 
O} ({\bar \psi},{\bar \xi})$. Using Eqn.~\ref{Tscale} we can rewrite
the momenergy deposition rate in terms of the disk temperature.
\begin{equation}
\label{Integral}
MDR = \alpha I(r_{\cal O},\theta_{\cal O})
\end{equation}
where $r_{\cal O}$ and $\theta_{\cal O}$ are the BL coordinate
location of the observer, the integral function is driven by the 
redshift of the neutrinos and anti-neutrinos arriving at the 
observer,  
\begin{equation}
\label{I}
I(r_{\cal O},\theta_{\cal O}) = \int\limits_\nu\int\limits_{\bar\nu} 
\left(
\frac{ {\bar \varepsilon}_{\cal O} }{ {\bar \varepsilon}_\parallel}
\right)^4 
\left(
\frac{ \varepsilon_{\cal O} }{\varepsilon_\parallel}
\right)^5   
(\Omnu \cdot \Ombnu - 1)^2 \Omnu\, d\Omega_\nu\, d\Omega_{\bar\nu} 
\end{equation}
and the constant $\alpha$ depends strongly on the temperature of the
disk as measured by the Keplerian observer.
\begin{eqnarray}
\label{alpha}
\alpha & = & \frac{7 \pi^3 \zeta(5) D G_F^2}{2 h^6}
\left(kT_{disk}\right)^9 \nonumber \\
& = & 1.84 \times 10^{32}\ \left(
\frac{kT_{disk}}{10 MeV}\right)^9\ \frac{ergs}{sec-cm^3}
\end{eqnarray}
  
\begin{figure}[ht] 
\begin{center}\epsfig{file=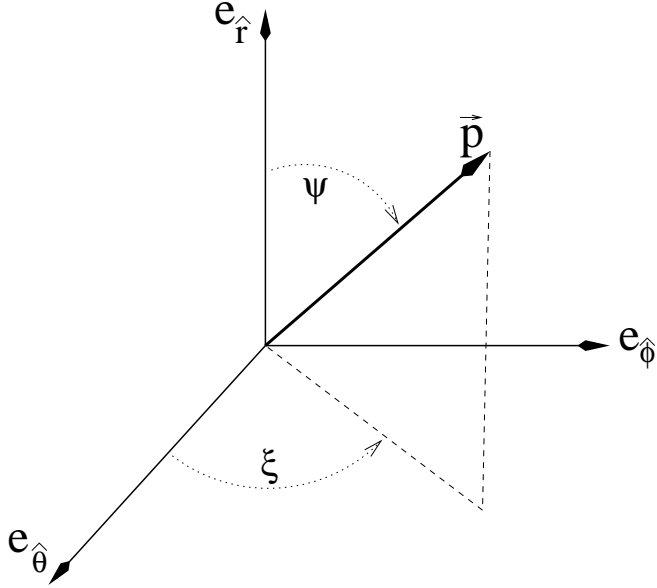,width=0.75\textwidth} \end{center} 
\caption{\label{skyangles} Spherical coordinate sky angles ($\psi$ and
$\xi$) used, in part, to generate the 4-momentum $\bpp$ in the
observers rest frame.}
\end{figure} 

In order to calculate the generation of 4-momentum from the 
$\nu$--$\bar \nu$ scattering, we must first image the accretion disk 
in the observer's reference frame. We do this by using a full 
3-dimensional ray-tracing code for a Kerr spacetime. The input needed 
by the ray-tracing code in order to image a given pixel of the disk is 
the location of the observer (${\cal O} = \{ t,r,\theta,\phi\}$) and 
the components of the 4-momentum ($\bpp$) in Boyer-Lindquist 
coordinates. 

\begin{eqnarray*} 
\bpp &=& p^{\hat 0} \be_{\hat 0} + p^{\hat r} \be_{\hat r} + 
         p^{\hat \theta}\be_{\hat \theta} + p^{\hat \phi} \be_{\hat \phi} \\ 
     &=& p^{0} \be_{0} + p^{r} \be_{r} + p^{\theta} \be_{\theta} + 
         p^{\phi} \be_{\phi} 
\end{eqnarray*} 
We can express the 4-momentum of the neutrino in the observer's frame 
by using the two spherical sky angles ($\psi$ and $\xi$) shown in 
Fig.~\ref{skyangles}. 
\begin{eqnarray*} 
p^{\hat 0}      &=& \vert\vec p\vert \ =\  \epnu \\ 
p^{\hat r}      &=& \vert\vec p\vert \cos{(\psi)} \\ 
p^{\hat \theta} &=& \vert\vec p\vert \sin{(\psi)} \cos{(\xi)} \\ 
p^{\hat \phi}   &=& \vert\vec p\vert \sin{(\psi)} \sin{(\xi)}   
\end{eqnarray*} 
The magnitude of the 3-momentum ($\vert\vec p\vert$), and the 
time component of the 4-momentum, were obtained by assigning an energy 
to the neutrino in the observers frame ($\epnu$) and using the 
normalization condition, $\bpp \cdot \bpp = 0$.  We can 
recover these components in the Boyer-Lindquist coordinates by 
inverting Eqns.(\ref{e0h},\ref{erh},\ref{eth},\ref{eph}). 
\begin{eqnarray*} 
p^{t}      &=& \frac{ \epnu }{ \sqrt{-g_{tt} }} - 
               \sqrt{ \frac{g_{tt}}{g_{\phi\phi} g_{tt} - 
               g_{t\phi}^2}}\ 
               \left(\frac{ g_{t\phi} }{ g_{tt} }\right)\  \epnu \sin{(\psi)} 
               \sin{(\xi)} \\ 
p^{r}      &=& \frac{ \epnu }{ \sqrt{g_{rr}} } \cos{(\psi)} \\ 
p^{\theta} &=& \frac{\epnu}{\sqrt{g_{\theta\theta}}} 
               \sin{(\psi)} \cos{(\xi)} \\ 
p^{\phi}   &=& \sqrt{ \frac{ g_{tt} }{ g_{\phi\phi} g_{tt} - g_{t\phi}^2} } 
               \ \epnu \sin{(\psi)} \sin{(\xi)}   
\end{eqnarray*} 

To reconstruct the image of the entire disk at ${\cal O}$, we 
integrate over all values of the sky angles ($\psi \in \{0,\pi\}$ and 
$\xi \in \{0,2\pi\}$). We use the ray-tracing code\cite{BroCheMil1997} to 
calculate the 4-momentum of the neutrino ($\bpp_{\|}$)when it hits the 
disk. 

\begin{figure} 
  \begin{center} 
  \epsfig{file=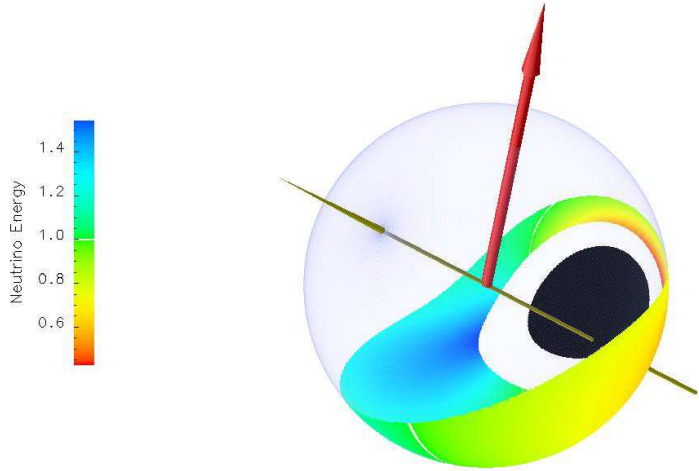,width=1.23\textwidth} \end{center} 
  \caption{\label{diskimage} A color rendering of the sky map of an observer located 
  above a rotating ($a=0.5$) black hole of mass ($M$). The observer is 
  located a distance $8M$ from black hole at an inclination 
  $\theta=3\pi/8$.  The thin Keplerian accretion disk extends from the 
  inner-most stable orbit ($r\approx 4.233M$) out to $r=10M$. We 
  assume that the neutrinos are emitted from an isothermal disk of 
  temperature $T_{disk}=1$ (as measured by the local comoving 
  Keplerian observers). The spectrum of the neutrinos and 
  anti-neutrinos are assumed to be a Planck spectrum characterized by 
  this temperature. The long thin yellow-colored coordinate axis 
  represents the observer's orthonormal ${\hat r}$ coordinate 
  axis. The thicker red arrow originating from the center of the 
  observer's celestial sphere represents the sum of all the scattered 
  momenergy as projected on the observer's spatial frame.  The image 
  of the accretion disk is shown color coded by the energy of the 
  arriving neutrinos as measured in the observer's frame. In 
  particular, the thin white, saturated band on the disk image 
  represents zero frequency shift. The neutrinos from the blue region 
  (darker region on the back of the sphere) are blue shifted due to 
  the closing relative velocity of the observer and the rotating disk 
  material, and the reddened regions indicate the gravitational red 
  shift effect from the black hole.  The non-axial symmetry of the 
  disk image, black hole and region between the horizon and the inner 
  edge of the disk is due to the black hole's rotation 
  (i.e. Lense-Thirring effect, frame dragging or gravitomagnetism). 
  This image was generated using a grid on the sphere of dimensions 
  $800\times1600$.} 
\end{figure}

\begin{figure} 
  \begin{center} \epsfig{file=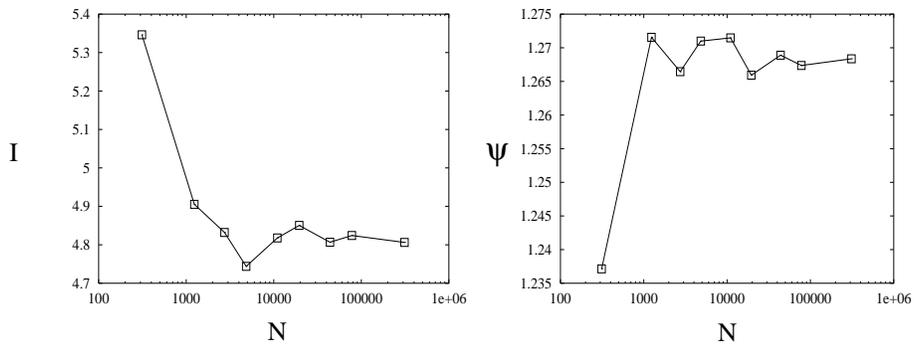,width=1.0\textwidth} 
  \end{center} 
\caption{\label{convMDR} A demonstration of the convergence of the
  energy deposition rate and tilt of the scattered electrons/positrons
  as a function of the number of pixels used to image the accretion
  disk. In this particular convergence test, we use an observer
  located at $r=8M$ and $\theta = 7\pi/16$ in an ($a=0$) Schwarzschild
  spacetime. We chose the location of this observer to correspond to
  the region of maximal momenergy production (as shown in the next
  figure). We plot the scattering integral $I$ in the left plot, and
  the angle $\psi$ in the right plot as a function of the number of
  grid points used to resolve the accretion disk.  The maximum
  resolution used in this convergence test involved tracing
  $2\times10^6$ geodesics over the entire celestial sphere. Of these
  geodesics, approximately one in seven, or $3\times10^5$, hit the
  disk.}
\end{figure}

In Fig.~\ref{diskimage} we provide an illustrative example of the 
imaging of an accretion disk extending from $r\approx 4.233M$ (the 
inner-most stable orbit of a Kerr black hole of mass $M$ and specific 
angular momentum $a=0.5$) out to $r=10M$.  Here the observer ${\cal O}$ 
is located at BL coordinates $r=8M$ and $\theta=3\pi/8$. The image of the 
disk is color coded as to the energy of the neutrinos reaching the observer. 
This image is used to calculate the MDR (Eqn.~\ref{MDR2}).  We also provide 
a convergence plot for the MDR in Fig.~\ref{convMDR}. 

\begin{figure} 
  \begin{center} \epsfig{file=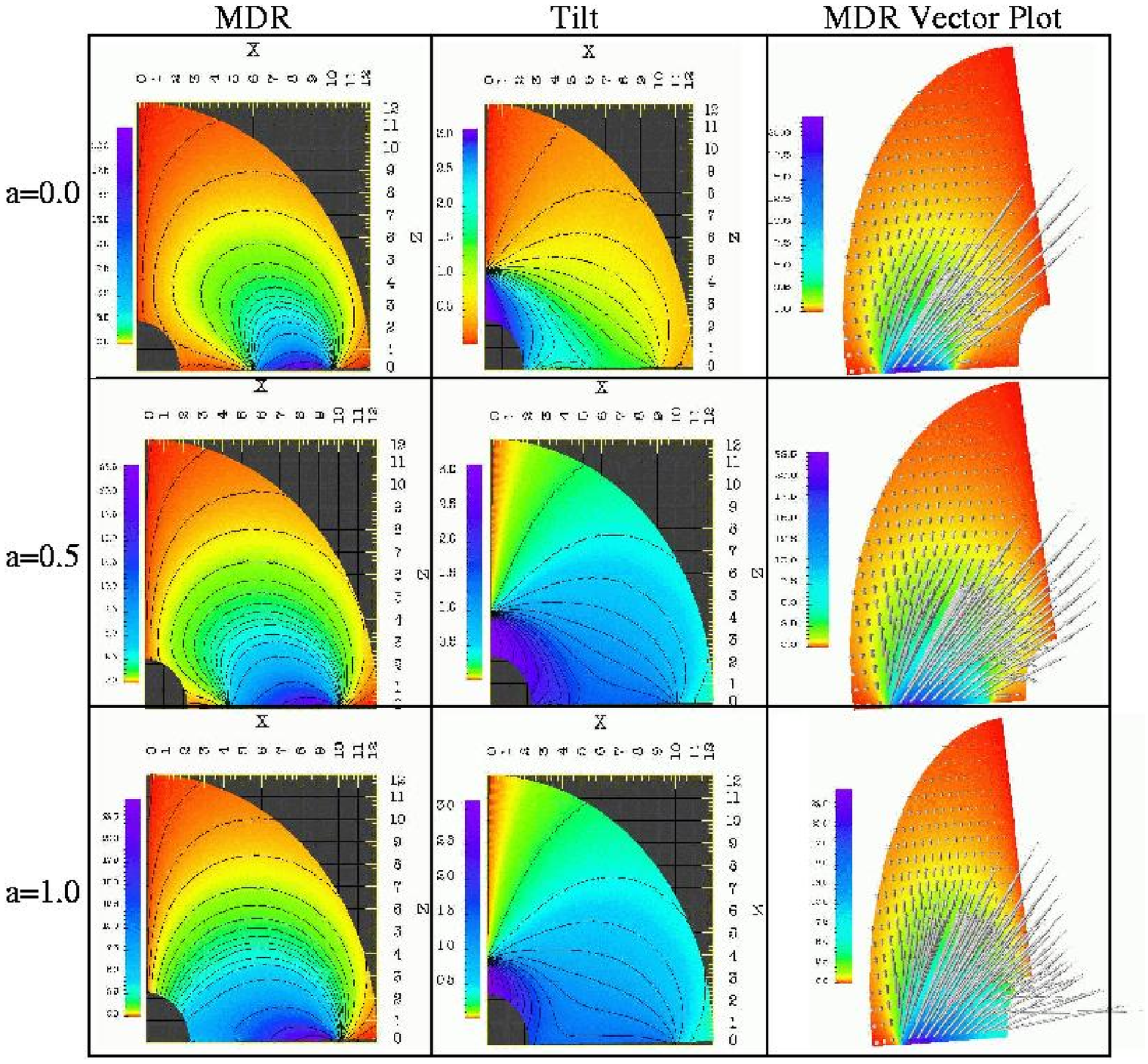,width=1.0\textwidth}
  \end{center} \caption{\label{pictures} Pictorial representation of
    our simulations for three values of the spin of the black hole.
    The first row represents a Schwarzschild black hole ($a=0$), while
    the third row an extreme Kerr hole ($a=1$). The second row
    represents our results for a moderately rotating black hole
    ($a=0.5$). The accretion disk in each of these cases extend from
    $10M$ in toward the black hole down to the inner-most stable
    orbit. The first column shows a contour plot of the magnitude of
    the time component of the momenergy deposition rate in Kerr
    coordinates, the second column depicts a contour plot of the
    angle, or tilt, the spatial components (in Boyer-Lindquist
    coordinates) of the MDR make with the symmetry $z$-axis. In the
    third column, we combine the information in the first two columns
    and pictorially show the spatial Boyer-Lindquist components of the
    momenergy deposition rate for various locations about the black
    hole accretion disk system.  One can observe the dominant
    contribution to the MDR above the surface of the disk produced by
    the $\nu \bar{\nu}$ collisions as well as their pitch angle.
    Values of the MDR can be obtained by multiplying the numbers in
    the graph by $\alpha$ and by the cube of the black hole's mass.}
\end{figure}

\section{Estimate  of the Transport of the Momenergy out to 
the Jet-Producing Region} 
\label{sec5} 

Up to this point we have calculated the total momenergy density
created in a given observer's orthonormal frame per unit 3-volume per
unit proper time.  These results are posed properly in general
relativity as they represent a meaningful local observation. However,
for the GRB application we address in this paper it would be useful to
determine the electron distribution function for an observer at large
radii (e.g. $r=50M$) from the accretion disk. This location would
approach the inner envelope of a hypothetical stellar shell
surrounding the black-hole/accretion disk system.  At such a location
we could construct a source term for the fluid at the inner envelope
of the star that would provide the energy for a jet which would
ultimately punch through the outer atmosphere of the star -- a jet
that would be the generator of a fireball-type solution.
Unfortunately, the construction of such a source is a computationally
difficult problem even under the ideal assumptions that the electrons
created evolve as a collisionless fluid and are highly energetic ($E
\gg m_e$). If we want to know how many electrons with momentum $p$
arrive at an observer located at spacetime event $x$, then we would
need to construct the geodesic into the past tangent to $p$ at $x$ and
sum up the electron creation functions along and tangent to that
geodesic.\cite{Ger2001} This would require modifying the calculation
of the MDR (Eq.~\ref{MDR}) to calculate the probability distribution
functions for the creation of electrons at each point.  We would then
need to construct and store these electron creation functions on a
dense grid in the $z$-$x$ plane perpendicular to the disk and
integrate these creation distribution functions over the 3-parameter
(2 sky angles and electron energy defining $p$) family of geodesics of
the observer at $x$.  Alternatively, we could solve a 3-dimensional
collisionless Boltzmann equation for the Kerr spacetime
geometry.\cite{LieMezThi2001} Unfortunately, both of these approaches,
while correct in a GR sense, are computationally demanding in the
extreme.  While there is no technical reason that would prevent us
from carrying out these calculations, practical considerations prevent
us from completing them in the near future. Consequently, we elect to
pursue a simpler computational strategy for the estimation of the GRB
jet source term.

 The difficulties with a general covariant treatment originate with
the Kerr black hole and not with the Schwarzschild solutions. In
particular, we currently calculate the total 4-momentum deposited per
unit 4-volume in the observer's orthonormal frame. Any attempt to
assign a 4-volume to the observer in BL coordinates (or any
coordinate) would break general covariance. In other words, the
spatial slices of the BL metric are not orthogonal to the time lines
($g_{\phi t} \ne 0$), and thus the 3-space of our observers is not
integrable.  Nevertheless, if we take the total MDR calculated in the
last section and transform this to the BL coordinates and multiply by
the ``effective BL spatial 3-volume'' of our observer, we can obtain
an approximation to a 4-momentum per proper time of electrons created
at that location. This is a coordinate-dependent quantity which should
be adequate for a qualitative understanding as long as the observers
are not too close the the horizon of a rotating black hole.  Moreover,
electron-positron production close to the horizon should not be
expected to contribute significantly to the results at large radii, as
the electron-positron production for observers close to the horizon
will be dominated by in-scatter of the electrons and positrons into
the black hole.  Given coordinate dependent MDR quantities, we then
parallel transport each of these 4-dimensional momenergy vectors
parallel to themselves out to the inner shell of the collapsar (say at
$r=50M$).  We then determine the time dilation and bin the results as
a function of the poloidal angle.  In this way we can estimate the MDR
from neutrino scattering as a function of the angular momentum of the
black hole.

This treatment is obviously non-covariant. However, as the
astrophysical uncertainties of the environs of such a black
hole/accretion disk system appear to be, at present, far greater than
the issues with relativity we raise here; we feel that our violation
of general covariance to simplify the computational requirements may
be justified in order to provide at least a qualitative understanding
of the off-axis energy-deposition at the inner envelope.  We
acknowledge that for a more complete characterization of the source
term, a covariant treatment should ultimately be undertaken. In the
meantime, we feel that the results presented herein provide a useful
qualitative estimate.

We have provided a pictorial representation of the total MDR
calculated for a field of observers in the $z$-$x$ plane perpendicular
to the accretion disk in Fig~\ref{pictures}.  Here we have placed
$10^4$ observers in this plane and calculated the MDR for each using
Eq.~(\ref{MDR2}).  The observers were evenly spaced in 100 equal BL
increments ($\Delta r$) in radius from $r=2.5M$ out to $r=12M$, and
equally spaced in 100 BL poloidal increments ($\Delta \theta$) from
$\theta=0$ to $\theta=\pi/2$. The MDR densities were transformed to BL
coordinates.  The 3-volume assigned to each observer is, as mentioned
above, coordinate dependent ($\Delta V= 2 \pi \sqrt{g_{ij}} \Delta r
\Delta \theta$) The last column of the figure illustrates the
structure of the MDR generated above the surface of the accretion
disk.

\begin{figure} 
  \begin{center} \epsfig{file=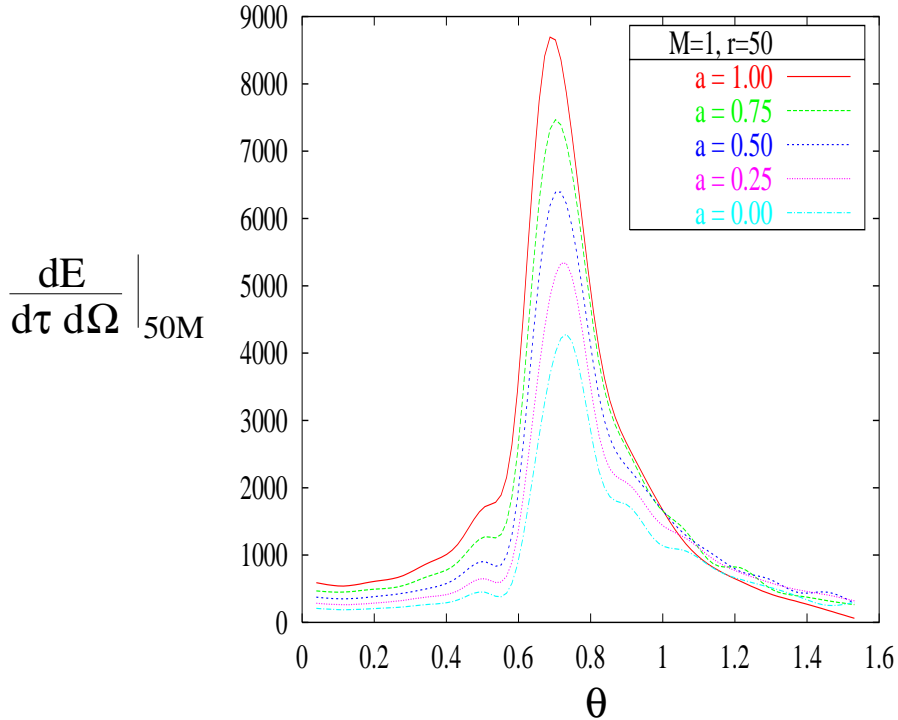,width=1.0\textwidth}
  \end{center} \caption{\label{angprofiles} We bin the MDR per unit
    solid angle as a function of the poloidal angle ($\theta$) at
    $r=50M$ for black hole/accretion disk systems with five values if
    the specific angular momentum ($a=0,.25,.5,.75\ \mbox{and}\ 1$).
    Although $dE/d\tau d\Omega$ levels off and yields results
    consistent with the conjecture of AF that the MDR is constant for
    small angles away from the axis; nevertheless, we do find a
    substantial disk-driven wind source off-axis. We demonstrate here
    the following two features: (1) The largest contribution to the
    MDR originates just above the disk giving factors of 10 to 20
    above the on-axis non-relativistic contributions calculated by
    others.  This gives rise to a disk-driven wind.  This wind, when
    propagated out to $r=50M$ along geodesics produces the peak in the
    MDR profiles shown in the figure. This peak corresponds to
    depositing the bulk of the MDR in a cone with opening angle
    $\theta \approx \pi/4$; and (2) The MDR only mildly depends on the
    specific angular momentum of the disk and in an approximately
    linear fashion.  The excess MDR for the extreme Kerr system
    results from the size of the accretion disk. In the Kerr system
    the disk extends from just above $r=M$ out to $r=10M$; however, in
    the Schwarzschild case the disk extends from $r=6M$ out to
    $r=10M$. These runs were performed using $10^4$ observers located
    at constant $\delta r$ and $\delta \theta$ intervals in the
    $z$-$x$ plane. The observers were placed between $r=2.5M$ and
    $r=10M$.  The MDR integral was calculated at each of these
    observers using $2\times10^4$ grid points covering the celestial
    sphere. Values of the MDR can be obtained by multiplying the
    numbers in the graph by $\alpha$ and by the cube of the black
    hole's mass. }
\end{figure}

\begin{figure} 
  \begin{center} \epsfig{file=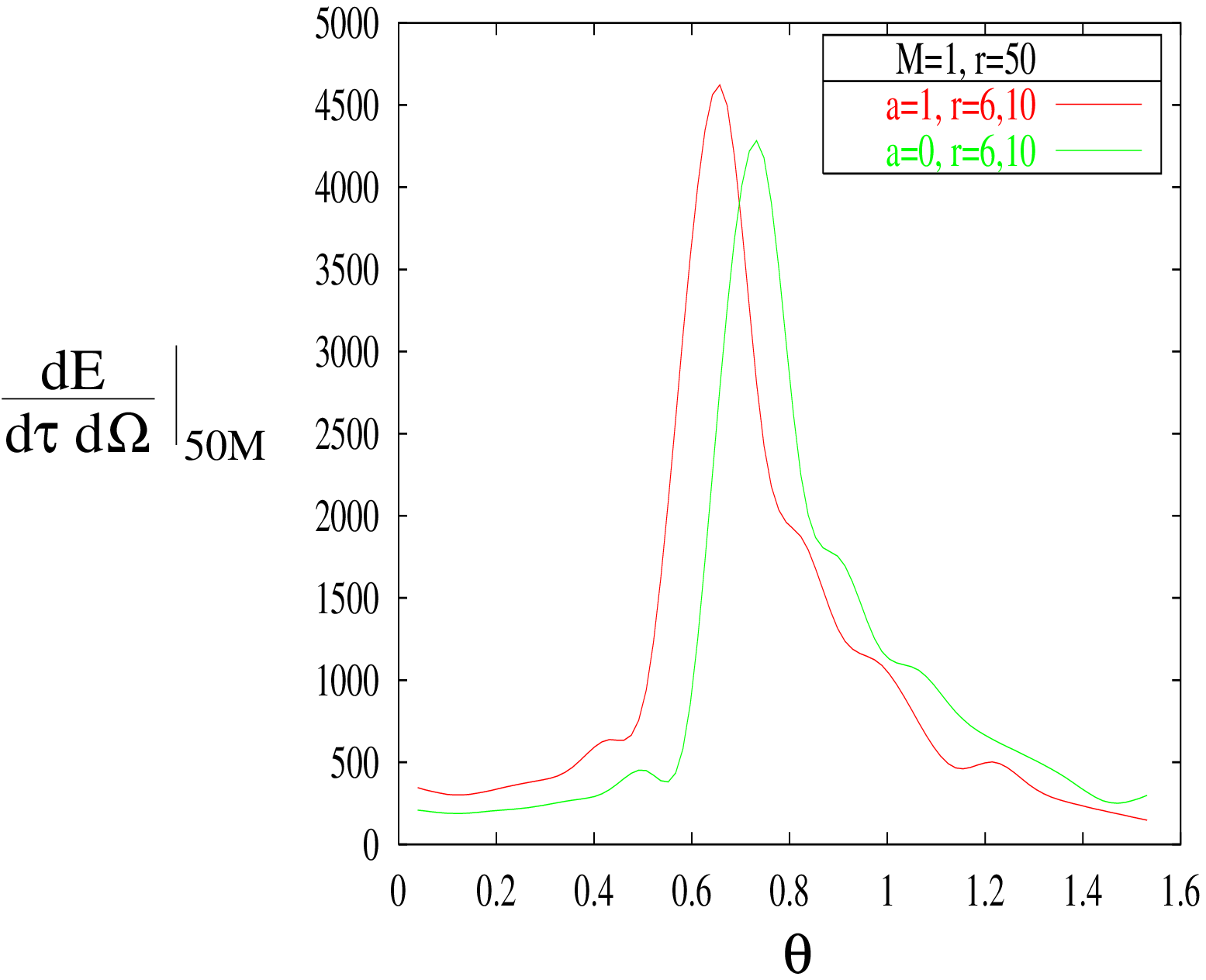,width=1.0\textwidth}
  \end{center} \caption{\label{erd_on_a} Demonstration that the MDR is
    primarily driven by the size of the disk and thus the location of
    the inner-most active region (active in the sense of neutrino
    production) of the accretion disk.  In the previous figure
    (Fig.~\ref{angprofiles}) we saw a factor of two increase in the
    peak MDR for an extreme Kerr black hole as compared to the
    Schwarzschild system. This is because the disk around the Kerr
    black hole extended down to the horizon, while in the
    Schwarzschild spacetime the disk ended at the inner most stable
    orbit at $r=6M$. In this figure we compare these two systems with
    the same disk. In both simulations ($a=0$ and $a=1$) we use an
    isothermal accretion disk with inner and outer radii of $6M$ and
    $10M$, respectively.  However, one simulation uses an extreme Kerr
    black hole ($a=1$), while the second simulation uses a
    Schwarzschild hole ($a=0$). The effect of angular momentum, sans
    location of the inner most stable orbit, has a marginal effect on
    both the magnitude of the total momenergy deposited at $50M$ and
    on the direction of the momenergy.  Values of the MDR can be
    obtained by multiplying the numbers in the graph by $\alpha$ and
    by the cube of the black hole's mass.}
\end{figure}

In Fig.~\ref{angprofiles} we integrated the BL MDR 4-momentum out
along their geodesics until they either hit the black hole or they
reach $r=50M$. We do this for five black hole/disk systems. The five
systems correspond to specific angular momentums with the values,
$a=0.0,0.25,0.5,0.75\ \mbox{and}\ 1.0$. For each value of $a$ the BL
time component of the MDR in the BL frame reaching $r=50M$ was binned
into 20 bins ranging from $\theta=0$ to $\theta=\pi/2$. The relative
magnitude of these curves increase with increasing values of the
specific angular momentum of the central black hole.  We have
investigated the origin of this increase and find that the source of
the increase is the region of the accretion disk near the inner-most
stable orbit. This is illustrated in (Fig.~\ref{erd_on_a}) where we
compare the angular MDR profile for an accretion disk extending from
$6M$ out to $10M$ with (1) a Schwarzschild black hole and (2) and
extreme Kerr black hole. The magnitudes are approximately equal and
only a slight tightening of the angle of the peak MDR is
apparent for the extreme Kerr black hole system.

\begin{figure} 
  \begin{center} \epsfig{file=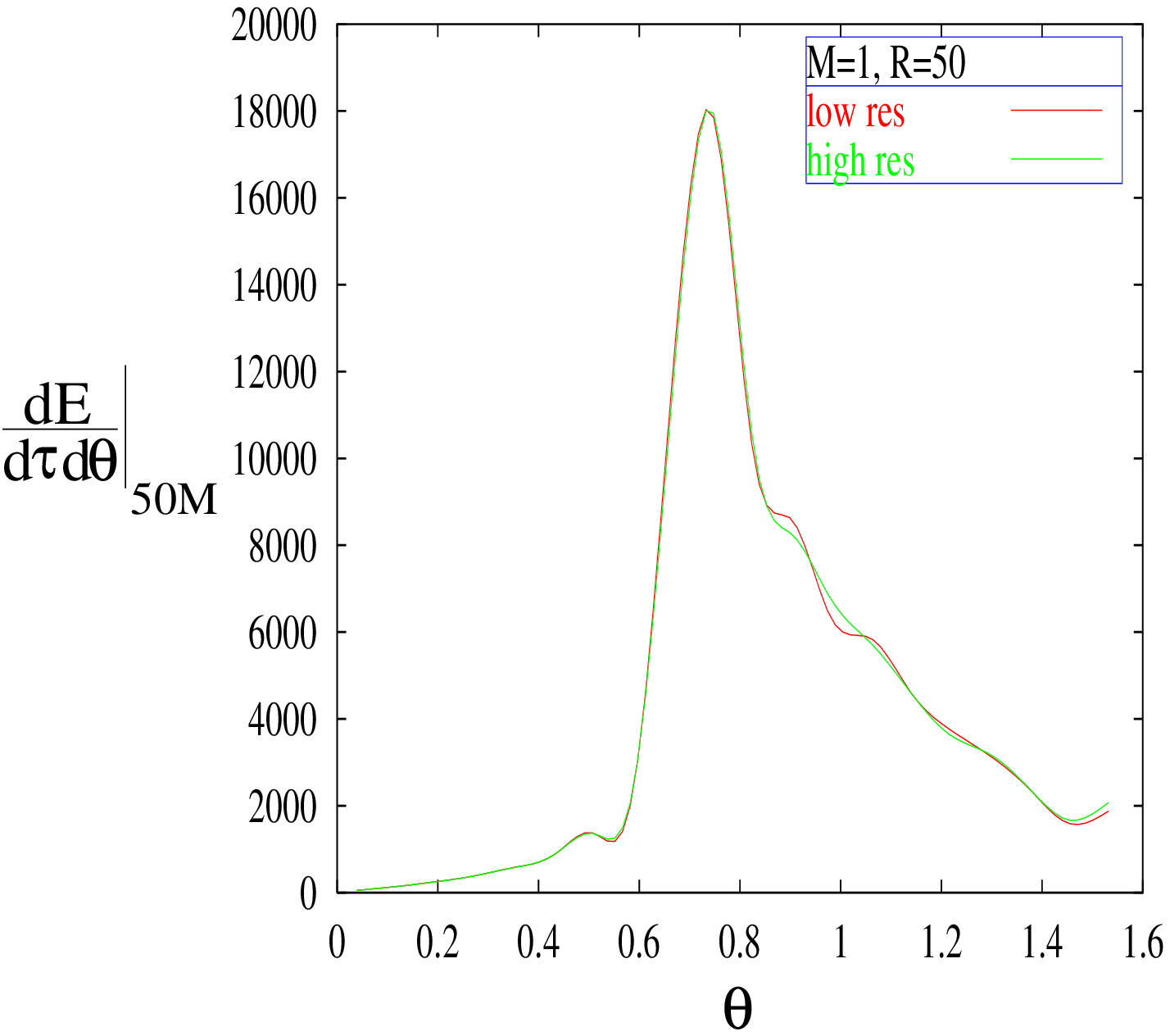,width=1.0\textwidth}
  \end{center} \caption{\label{highlow} Demonstration of the
    convergence of our calculations of the MDR at $50M$ using
    different number of observers located in the $z$-$x$ plane around
    an $M=1$ Schwarzschild black hole/accretion disk system.  The high
    resolution curve uses $1.5\times10^4$ observers in the $z-x$
    quadrant(equally spaced between $r=2.5M$ and $r=12M$), while the
    low resolution curve uses $1\times 10^4$ observers in this
    quadrant. In both cases the MDR scattering integral used
    $100\times 200$ grid points. Here we plot the MDR per unit
    $\theta$ as opposed to MDR per unit $\Omega$ in the other figures.
    Values of the MDR can be obtained by multiplying the numbers in
    the graph by $\alpha$ and by the cube of the black hole's mass.}
\end{figure}

We demonstrate convergence of our results with regard to the number of 
observers we use in the $z$-$x$ plane by comparison of a high and 
lower resolution run (Fig.~\ref{highlow}). The higher resolution, 
which we use in our calculations in this manuscript, is a little smoother 
than the lower resolution run; however, the both compare well with each 
other. 

\begin{figure} 
  \begin{center} \epsfig{file=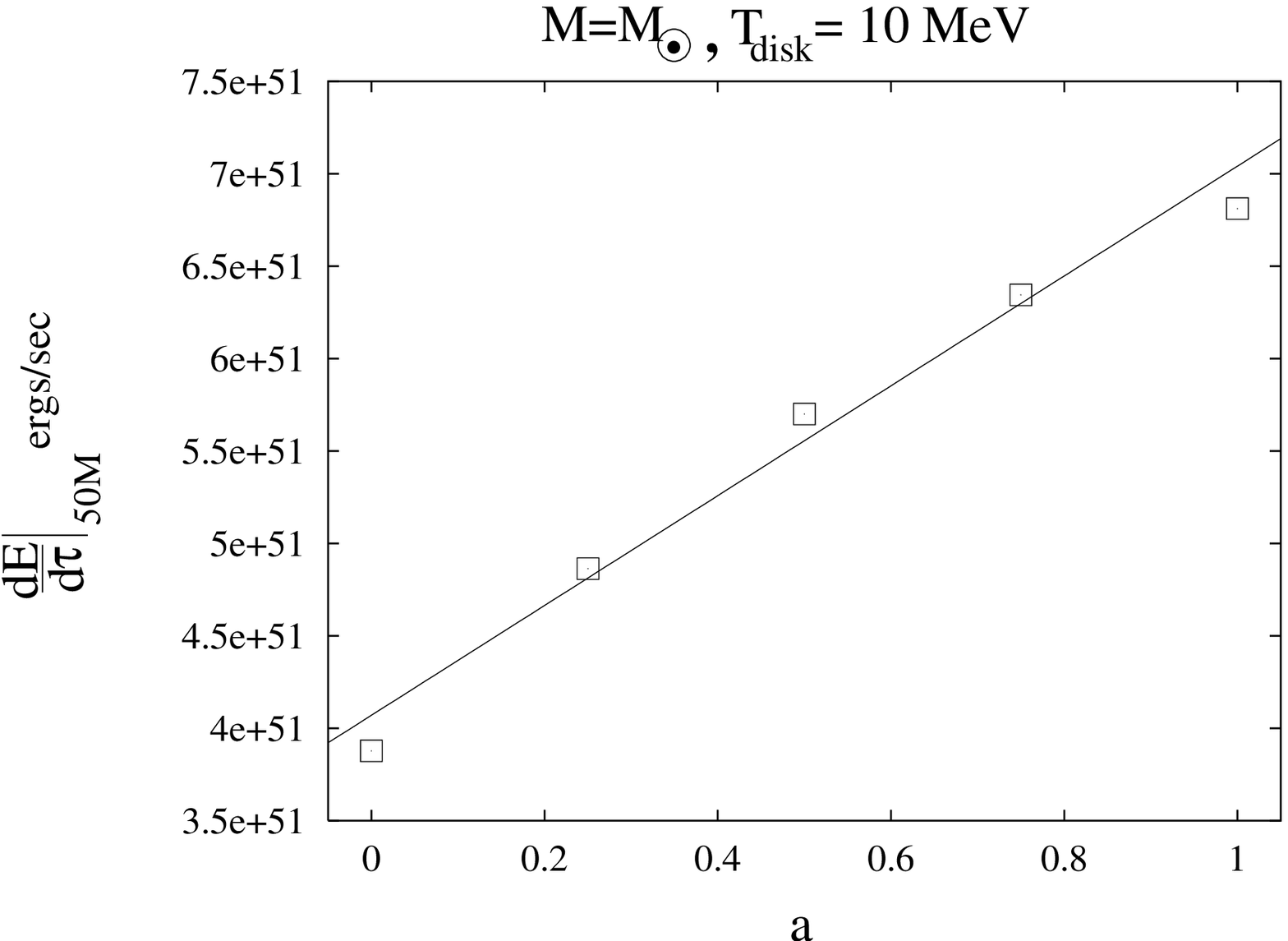,width=1.0\textwidth}
  \end{center} \caption{\label{totaledr(a)} We plot the total BL time
    component of the MDR at $50M$ as a function of five values of the
    specific angular momentum ($a$) of a solar-mass black hole with a
    10 MeV temperature accretion disk. In other words, we integrate
    each of the five profiles in Fig.~\ref{angprofiles} over the solid
    angle $\Omega$. This shows that the MDR is roughly a linear
    function with respect to the specific angular momentum ($a$) of
    the black hole.  We provide a least-squares fit of the five
    numerical points.  The integration of the MDR profiles are done
    over the entire northern hemisphere. If we integrated our profiles
    only over a small solid angle around the symmetry axis we would
    find our result are consistent with those reported by AF. However,
    we include here the disk-driven wind created by $\nu$--$\bar{\nu}$
    scattering above the disk's surface. We see a large factor
    (approximately 10-20 times) in the MDR from the disk surface over
    and above the on-axis source calculated by AF and SW. This
    enhancement occurs equally well for Schwarzschild black holes as
    for Kerr black holes.}
\end{figure}

The relevant quantity reported by AF was the total MDR at a distant
observer. If we integrate each of the five MDR curves shown in
Fig.~\ref{angprofiles} over $\Omega$ at $r=50M$ we will obtain the
integrated MDR into the jet as shown in Fig.~\ref{totaledr(a)}. It
will be concentrated along the axis and peak along a cone of opening
angle $\pi/4$.  This opening angle is due to the tilt of the MDR near
the surface of the disk. The tilt is in the direction of the disk's
rotation.  We find a linear dependence for this integrated MDR as a
function of the specific angular momentum of the black hole. This is
an extension of the earlier results because we are including the
relatively large off-axis source term -- the MDR created just above
the surface of the disk via $\nu$ $\bar{\nu}$ scattering.  This disk
MDR produces a 10 to 20 fold enhancement to the total MDR energy
deposited at $r=50M$ over and above the on-axis MDR, even for a
Schwarzschild black hole. Such a disk-generated MDR was also found in
the non-gravitational models of AF.\cite{AsaFuk2000} The dependence of
this disk-driven source of momenergy on the specific angular momentum
of the black hole is roughly linear.  We predict a twofold increase in
the MDR for an extreme Kerr black hole relative to a Schwarzschild
hole.  In addition, in (Fig.~\ref{angprofiles}) we only see a slight
tightening of the cone of the momenergy deposition with increasing
$a$.  It would be interesting to observe the jet dynamics through the
outer envelope of the star with sources of this kind.

\section{Conclusions: Enhancement of the Energy Deposition from a Disk-Driven Wind} 
\label{sec6}

The goal of the research in the GRB central engine modeling community
over the last few years was to look for a needed enhancement of a
factor of 10 in the MDR. Previous work by AF and SW looked for such an
enhancement over the non-gravitating models by examining the GR
effects on neutrino pair scattering along the symmetry axis.  They did
not find this enhancement; however, in this paper we identify the
off-axis disk-driven MDR component that yields such an increase over
corresponding on-axis MDR values. Furthermore, this disk-driven
momenergy occurs in the non-rotating black holes as well as the
rotating black holes, and is even present in the non-physical models
with no gravity effects (e.g. Fig 4 of Ref.~\cite{AsaFuk2000}). We
therefore have strong indication that the the current fireball central
engine model discussed here is viable. Of course, more detailed
astrophysical simulations are needed which include the complex
environs of the black hole disk system, geometry and transient
behavior of the disk and black hole, and a simulation of the induced
plasma transport of the scattering out to the fireball region.  Since
this disk-driven MDR source is relatively large, and since the baryon
pollution problem\cite{Pac1990} remains an open issue in this
field,\cite{Mez2002} we argue that it may be premature to discount
this off-axis source. We should reserve such a decision until more
detailed GR, hydrodynamic, MHD and plasma transport calculations have
been done.
 
As we outlined in the paper, we have extended the earlier work of AF
to calculate the off-axis contribution to the MDR. When using the
assumptions made by these authors, our results are consistent their
work; however, we find the generation of a substantial MDR above the
disk, which, when added to the on-axis calculations of these authors
produce a substantial enhancement to the deposition of
energy-momentum. In particular, we find an approximately linear
dependence of the total MDR at large $r$ as a function of specific
angular momentum of the black hole.
\begin{equation}
\left. \frac{dE}{d\tau}\right|_{50M} \approx 
4.06 \times 10^{51}\ \left( 1+.73 a \right)
\ \left(\frac{kT_{disk}}{10 MeV}\right)^9 
\left(\frac{M}{M_\odot}\right)^3 \ \frac{ergs}{sec} 
\end{equation}
Here we did a least-squares fit to the numerical results in
Fig.~\ref{totaledr(a)}, multiplied by $\alpha$ times the cube of the
solar mass ($M_\odot = 1.47\times 10^5 cm$).  The bulk of the MDR at
large $r$ is deposited in a conic region centered on the symmetry axis
of the system with an opening angle of approximately $\pi/4$.

Although we provide a rigorous general relativistic calculation of the
MDR at an arbitrary observer, computational constraints required us to
make some approximations regarding the transport of the scattered
energy out to the inner envelope ($r=50M$) of the star. Assuming (1)
the electrons produced have substantially higher energy than their
rest mass, (2) that the electrons propagate away in a collisionless
fashion and (3) that the observationally-relevant electron creation
distribution functions are peaked around the total MDR calculated,
then our results should be rigorously correct for a Schwarzschild
black hole, and qualitatively correct for spinning black holes. We
would not be surprised if a general-relativistically rigorous
calculations would yield different quantitative results for the
extreme Kerr black hole close to the horizon.  However, one of the main
results of this paper is the identification of the off-axis
disk-driven MDR as providing an enhancement to the energy deposition
rate rather than the GR-driven  enhancements.

We outlined in the last section an approach to calculate the electron
distribution function at an observer $\cal O$ at large $r$ in a
relativistically covariant way. This involved integrating the
electron-creation distribution functions along geodesics into the past
which originated at $\cal O$. One step in this direction would be to
extend the work here and calculate the electron creation distributions
in the frame of our observers as opposed to the total MDR. This is not
difficult in principle but is computationally demanding as it would
require evaluating a 6-dimensional scattering integral as opposed to
the 4-dimensional integral we evaluated numerically in
Sec.~\ref{sec4}, Eqn.~\ref{MDR2}.  The utility of such a calculation
would be to evaluate the point along the axis on the border between
out-scatter and in-scatter (i.e. when the total MDR was zero). Would
the electron creation function at and about this event yield a
non-negligible contribution to the electron distribution function at
large $r$?

On a more speculative note, we also are interested in calculating
secondary scattering of the electrons and positrons within the
ergosphere of the black hole. We have found evidence of significant
in-scatter into this region. Will the Penrose process produce some
observable high energy events from this region that can give us more
clues to the inner workings of the central engine of these low mass
black hole/accretion disk systems?  

One question seems of paramount importance in this field: {\sl What
  observational features of a GRB can give clues as to the
  inner-workings of the central engine?}\cite{Woo1999} Surely future
X-ray, $\gamma$-ray and gravity wave observatories complemented with
theory can soon unveil these cosmic explosions.

\section{Acknowledgments} 
We wish to thank Doug Eardley, Adrian Gentle, Bob Geroch, Daniel Holz
and Rosalba Perna for stimulating conversations on this subject, and
Chris Fryer for suggesting this problem. We thank the Institute for
Theoretical Physics at UC Santa Barbara for providing a supportive and
stimulating research environment to complete this research. This
project was supported by Los Alamos under the LDRD/ER program and in
part by the National Science Foundation under Grant No. PHY99-07949.

\end{document}